# Real-time 112 Gbit/s DMT for Data Center Interconnects

A. Dochhan[1], N. Eiselt[1], J. Zou[1], H. Griesser[2], M. H. Eiselt[1], J.-P. Elbers[2]
*1 ADVA Optical Networking SE, Maerzenquelle 1-3, 98617 Meiningen, Germany*
*2 ADVA Optical Networking SE, Fraunhoferstr. 9a, 82152 Martinsried, Germany*
*adochhan@advaoptical.com*

**Abstract:** We report on 112 Gbit/s real-time DMT transmission over up to 60 km, targeted at DCI applications. Chromatic dispersion mitigation by vestigial sideband filtering is compared to the use of dispersion compensating fiber.

## 1. Introduction

Data center interconnects (DCI) links typically span a distance between 10 to 80 km. Besides upcoming low-cost coherent technologies driven by Silicon photonics, IM/DD formats with ≥ 50 Gbit/s per wavelength are promising short-term solutions to enable multi-Tbit/s transmission with stringent cost, form factor and power consumption restrictions. While 4-level pulse amplitude modulation (PAM4) would enable the re-use of a short-reach standardized ecosystem [1], it exhibits very limited tolerance towards chromatic dispersion (CD), requiring either optical CD compensation [2] or advanced digital signal processing (DSP), such as vestigial or single side band (VSB/SSB) transmission with CD pre-compensation or reconstruction of the optical field [3-4]. In contrast to PAM4, discrete multi-tone transmission (DMT) shows a higher CD tolerance, enabled by its adaptive bit and power loading (BL/PL) capability, but highest capacities have also been demonstrated with advanced DSP, like SSB, VSB and interference cancellation [5]. Recently, 440 Gbit/s per λ have been shown using a Kramers-Kronig receiver and entropy loading [6]. However, real-time results with low-cost PHYs have not been demonstrated so far.

In this paper, we use a commercially available DMT DSP chip to generate and to transmit 112 Gbit/s per wavelength DMT signals over up to 60 km of standard single mode fiber (SSMF). No advanced DSP for subcarrier-subcarrier beating interference (SSBI) cancellation or for CD pre-compensation is applied. We compare the reach of double sideband (DSB) DMT, achieved with dispersion compensating fiber (DCF), to VSB DMT, enabled be the use of the DWDM multiplexer (MUX) and de-multiplexer (DEMUX) in the system.

## 2. Experimental Setup

The experimental set up is shown in Fig. 1. The DMT signal was fed differentially into a linear driver amplifier and sent to a $LiNbO_3$ Mach-Zehnder-Modulator (MZM) with a bandwidth of 27 GHz. The continuous wave laser at the MZM input was set to 193.4 GHz. For obtaining VSB signals, it was shifted by 33 GHz to use the 100-GHz MUX and DEMUX filters (~ 63 GHz bandwidth FWHM) for cutting of one sideband, resulting in 4.9 Tbit/s capacity when using the full C-band. The signal was boosted by an Erbium doped fiber amplifier (EDFA). After the transmission fiber, the signal was pre-amplified by another EDFA, which offers the option of placing a DCF mid-stage. Then, the signal was filtered by a DEMUX filter and detected by a PIN/TIA. The DMT PHY chip decoded the DMT signal and counted the bit error ratio (BER) in real-time. The chip features a forward error correction (FEC) code with code word length 9k and a nominal FEC limit of BER = 3.3e-3. The DMT signal consisted of 512 subcarriers, running at a sample rate of ~56 GSa/s. 16 samples were used as cyclic prefix. A clipping ratio of 12.5 dB was applied. For synchronization and framing, two consecutive pilot subcarriers are reserved, as can be seen from the estimated SNR and BL plots in Fig.1. Apart from DMT one-tap equalization no additional compensation scheme was applied.

## 3. Results and Discussion

Fig. 2 a) shows the estimated SNR with 100-GHz filters for DSB signals without DCF for b2b and for 10 and 40 km. Even for 10 km, the influence of CD on the signal is quite severe. DCF compensates for the CD induced notches, but introduces additional non-linearity. Fig. 2 c) shows the OSNR vs. BER results for transmission with DCF. The optimum fiber launch power was found to be 7 dBm, the launch power to the DCF was chosen 6 dB lower. Up to 60 km transmission were possible with a required OSNR of 43.5 dB at the FEC limit. For longer distances the system was mainly OSNR limited. The 1.5 dB transmission penalty compared to the b2b case can be explained by the influence of fiber non-linearity. To reduce the complexity of the setup, the MUX/DEMUX filters were used to produce a VSB signal and DCF was omitted. The optimum frequency shift of the laser was found to be 33 GHz. Fig 2 b) shows the estimated SNR vs. frequency for the b2b DSB signal, and VSB signals with 10 km and 40 km of SSMF. The influence of CD is reduced, but not mitigated completely as can also be seen from the BER vs. OSNR curves in Fig. 2 d). While there is no penalty for the VSB b2b signal compared to the DSB b2b signal, the required OSNRs for 10 km and 20 km are 45.5 and 47.5 dB, respectively. Besides incomplete CD mitigation, this effect is owed to the additional SSBI, which is not compensated by the DSP. Therefore, the required OSNR is much higher than suggested for DCI systems [7], but since most of the links are shorter than 80 km and the DSP allows a switch to half rate if the transmission quality is not good enough, DMT still might be a promising near-term solution for the DCI scenario. Moreover, PAM4 would show similar high requirements if implemented with efficient DSP and results are in good agreement with theory [8].

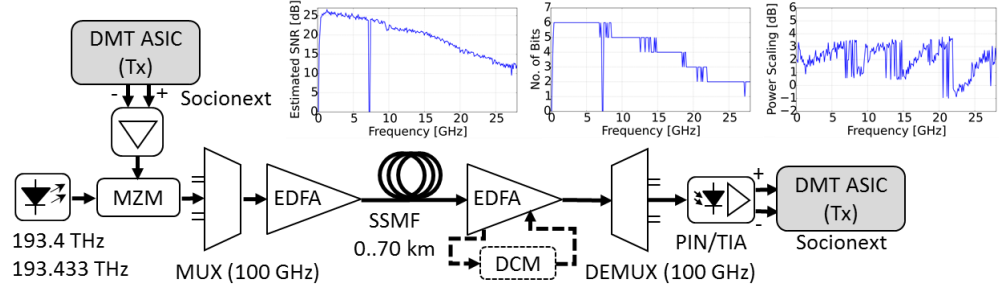

Fig. 1: Experimental setup for 112 Gbit/s real-time DMT transmission. Insets: Estimated SNR and bit and power loading for b2b.

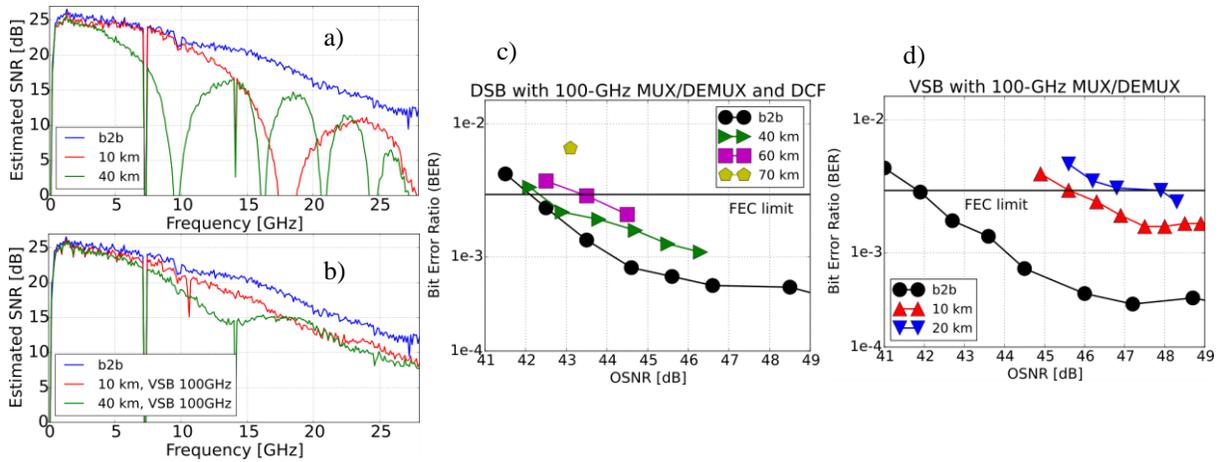

Fig. 2: a) Estimated SNR for b2b (blue), 10 km (red) and 40 km (green) transmission, without DCF, DSB signal; b) SNR for VSB signal; c) BER vs. OSNR for DSB signal with DCF; d) BER vs. OSNR for VSB signal using MUX/DEMUX as VSB filters.

### 4. Conclusion

We show successful real-time transmission of 112 Gbit/s DMT using a commercially available DMT PHY chip. Up to 60 km SSMF transmission was achieved using DCF and filters for 100-GHz channel spacing, while VSB filtering was shown to eliminate the need for DCF over distances up to 20 km. Since many of the targeted data center interconnects are shorter than the specified 80 km, DMT could be a viable solution, especially since the utilized chip offers the ability to switch to 56 Gbit/s per wavelength. Therefore, it can enable a flexible transmitter performing a trade-off between reach and data rate.

### 5. Acknowledgment

The authors would like to thank Socionext, especially Colin Brenton, for their support during the measurements. The results were obtained in the SENDATE Secure-DCI project, partly funded by the German ministry of education and research (BMBF) under contract 16KIS0477K, and in the iCirrus project, funded by the European Commission under grant agreement No. 644526.